\documentclass[twocolumn]{aastex631}

\shorttitle{Secondary electron spectrum in CR discharge}
\shortauthors{Kamiido Kazuki and Yutaka Ohira}
\usepackage{natbib}
\usepackage{lineno}
\begin{document}

\title{Collisionless shock in a relativistically hot unmagnetized electron-positron plasma}
\author[0009-0002-9063-8827]{Kazuki Kamiido}
\author[0000-0002-2387-0151]{Yutaka Ohira}
\affiliation{Department of Earth and Planetary Science, The University of Tokyo, \\
7-3-1 Hongo, Bunkyo-ku, Tokyo 113-0033, Japan}
\email{kamiido-kazuki8990@g.ecc.u-tokyo.ac.jp}

\begin{abstract}
In this work, we investigate collisionless shocks propagating in a relativistically hot unmagnetized electron-positron plasmas. 
We estimate the dissipation fraction at shocks in the relativistically hot plasma, showing that it is sufficiently large to explain the observation of gamma-ray bursts even when the shock is not highly relativistic. 
It is shown by two-dimensional particle in cell simulations that magnetic fields are generated around the shock front by the Weibel instability, as in the cold upstream plasma. 
However, in contrast to the cold upstream plasma, no particles are accelerated at the shock in the simulation time of $t = 3600~\omega_{\rm p}^{-1}$. 
The decay of the magnetic field in the downstream region is slower for slower shock velocities in the hot plasma cases. 
Applying the slow decay of the downstream magnetic field, we propose a model that generate magnetic fields in large downstream region, which is required from the standard model of the gamma-ray burst afterglow. 
\end{abstract}


\section{Introduction}
\label{sec1} 
Relativistic collisionless shocks have an important role in the dissipation of relativistic outflows from black holes and neutron stars. 
In the relativistic collisionless shock, magnetic field turbulence and high energy particles are generated, so that the kinetic energy or pointing flux of the relativistic outflow is expected to be efficiently converted to high-energy photons, cosmic rays, and neutrinos. 
Therefore, physics in the relativistic collisionless shock determines how brightly high-energy astronomical objects 
shine in electromagnetic waves and neutrinos. 
However, we have not fully understood the physics in the relativistic collisionless shock although there are many studies about the relativistic collisionless shock. 

Early studies showed via ab initio particle-in-cell (PIC) simulations that 
the synchrotron maser \citep{hoshino91} and Weibel \citep{weibel59} instabilities mainly dissipate the kinetic energy of upstream plasmas for magnetized and unmagnetized (or weakly magnetized) relativistic shocks, respectively \citep{langdon88,hoshino91,spitkovsky08a,spitkovsky08b}. 
In those simulations, the upstream plasma has a relativistic bulk velocity in the shock downstream rest frame, but the upstream temperature is a nonrelativistic temperature (less than the rest mass energy). 

There is another kind of relativistic shock that has a nonrelativistic or mildly relativistic upstream flow velocity in the shock downstream rest frame, but the upstream temperature is a relativistic temperature (larger than the rest mass energy). 
Recently, it has been shown that such a collisionless shock can be produced by the interaction of a main relativistic shock with a high-density clump \citep{inoue11,tomita22}.  
After the upstream high-density clump interacts with the main relativistic shock, the shocked clump pushes the downstream plasma of the main shock, so that another shock propagates in the relativistically hot downstream plasma. 
In addition, a fluid simulation with a high resolution showed the presence of such shocks in lobes of powerful radio galaxies \citep{matthews19}. 
So far, however, no detailed study of collisionless shocks in the relativistically hot plasma has been carried out using PIC simulations. 
We do not understand how the collisionless shock is generated in the relativistically hot plasma, whether a sufficient dissipation occurs to satisfy the Rankine-Hugoniot relation based on hydrodynamics, how long the shock transition region is, how magnetic field turbulence is generated in the shock transition region, whether particles are accelerated or not. 

We estimate the fraction of energy dissipated by shocks in the relativistically hot plasma in Section~\ref{sec2}, and 
present the first PIC simulation of collisionless shocks propagating in the relativistically hot plasma in Section~\ref{sec3}. 
We then propose a model that generate magnetic fields in large downstream region in Section~\ref{sec4}.
Section~\ref{sec5} is devoted to the discussion and summary.

\section{Dissipation fraction}
\label{sec2} 
\begin{figure}
\centering
\plotone{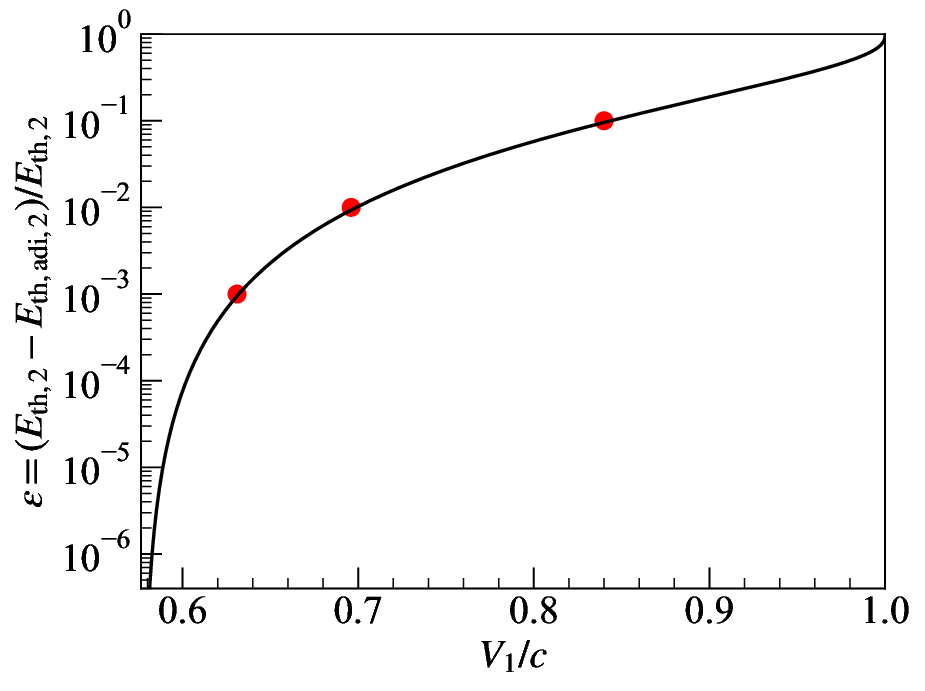}
\caption{Fraction of the energy dissipated by shocks in the relativistically hot plasma. 
The red points show the points where $ \varepsilon = 10^{-3}, 10^{-2}$, and $10^{-1}$.}
\label{fig1}
\end{figure}

In the downstream rest frame, the upstream energy flux is converted to the downstream thermal energy at the shock. 
The conversion process can be divided to the adiabatic and non-adiabatic processes. 
The adiabatic process is compression at the shock, where the entropy is conserved, but the non-adiabatic process generates the entropy. 
We here define the following fraction of the energy dissipated by the shock,
\begin{equation}
\varepsilon = \frac{E_{\rm th,2} - E_{\rm th,2,adi}}{E_{\rm th,2}} ,
\end{equation}
where $E_{\rm th,2}$ and $E_{\rm th,2,adi}= E_{\rm th,1}r^{4/3}$ are the downstream thermal energy 
and the downstream thermal energy generated by the adiabatic shock compression, 
and $E_{\rm th,1}$ and $r$ are the upstream thermal energy and compression ratio of the shock. 
The adiabatic index is set to be $4/3$ because we consider a relativistically hot plasma in the upstream region. 
From the Rankine-Hugoniot relation, $E_{\rm th,2}/E_{\rm th,1}$ and $r$ can be written by using the Mach number of $M=\sqrt{3}V_1/c$  \citep{kirk99}, where $V_1$ and $c$ are the shock velocity in the upstream rest frame and speed of light, 
\begin{equation}
\frac{E_{\rm th,2}}{E_{\rm th,1}} = \frac{3M^2-1}{3-M^2} , r = M \sqrt{\frac{3M^2-1}{3-M^2}}.
\end{equation}
Then, the dissipation fraction is
\begin{equation}
\varepsilon = 1 - 3\left [\frac{\{1-(V_1/c)^2\}(V_1/c)^4}{9(V_1/c)^2-1}\right]^{1/3}, 
\end{equation}
and shown in Figure~\ref{fig1}. 
Because the magnetic field generation and non-thermal particle acceleration around the shock are the non-adiabatic process, the dissipated energy fraction gives the upper limits of the downstream energy fraction of magnetic fields, $\varepsilon_{\rm B}$, and nonthermal electrons, $\varepsilon_{\rm e}$. 
Observed data about afterglows of gamma-ray bursts (GRBs) suggest $\varepsilon_{\rm B}\sim10^{-8} - 10^{-3}$ and $\varepsilon_{\rm e}\sim 10^{-1}$ in the downstream region of the forward shock propagating in interstellar or circumstellar media \citep{santana14}. 
Figure~\ref{fig1} shows $\varepsilon\sim 10^{-3}, 10^{-2}$, and $10^{-1}$ for $V_1 / c \sim 0.63, 0.69$, and $0.84$, respectively. 
Therefore, even for a mildly relativistic shock with a low Mach number, we can expect sufficient amounts of magnetic fields and nonthermal particles to explain the GRB afterglow. 
In addition, it should be noted that the dissipation fraction at the shock is almost unity at the relativistic shock limit ($V_1 / c \approx 1$) even when the the shock Lorentz factor ($\{1-(V_1 / c)^2 \}^{-1/2}$) is much smaller than the mean Lorentz factor of upstream plasmas ($\sim 3kT/mc^2$), where $k, T$ and $m$ are the Boltzmann constant, upstream temperature and particle mass, respectively.

\section{Simulation}
\label{sec3} 
\subsection{setting}
To investigate collisionless shocks propagating in a relativistically hot unmagnetized plasma, we perform two-dimensional PIC simulations using the open code, Wuming \citep{matsumoto24}, which solves the equations of motion for many charged particles and the Maxwell equations self-consistently
using the Buneman-Boris method for the equation of motion of particles, implicit FDTD scheme for the Maxwell equation \citep{hoshino13}, and Esirkepov's charge conservation scheme for the current deposit with the 2nd-order shape function \citep{esirkepov01}.
Particles with a bulk velocity in the $-x$ direction are continuously injected from the right boundary and reflected at the left wall at $x=0$. 
A periodic boundary condition is used in the y-direction for all particles and electromagnetic fields. 
As time goes on, a collisionless shock is formed in a self-consistent manner, and propagates in the $+x$ direction. 
Our simulation frame corresponds to the downstream rest frame.
The injected upstream plasma is set to relativistically hot electron-positron plasmas with the Maxwell-J\"{u}ttner distribution. The upstream temperature is set to $kT=5~mc^2$, where $k$ and $m$ are the Boltzmann constant and particle mass, respectively. 
There is no background magnetic field initially. 
The simulation box size is $L_x \times L_y =6200~c/\omega_{\rm p} \times 104~c/\omega_{\rm p}$, where $\omega_{\rm p}=(4\pi n_{\infty}e^2c^2/3kT)^{1/2}$ is the upstream electron plasma frequency for $kT\gg mc^2$, $n_{\infty}$ is the electron-positron density at the far upstream region. 
The simulation cell size and time step are set to $\Delta x=\Delta y=0.1~c/\omega_{\rm p}$ and $\Delta t = 0.1~\omega_{\rm p}^{-1}$, respectively.  
60 simulation particles per cell per species are injected in the upstream region. 
We perform several simulations with different bulk velocities, $V_{12}/c=0.1, 0.3, 0.5, 0.7$, and $0.9$. 
In addition, to clarify the effects of the relativistic temperature in the upstream region, we perform a simulation for a cold upstream plasma ($kT/mc^2 = 0.01, \Gamma_{12}=24.57$), where $\Gamma_{12}=\{1-(V_{12}/c)^2\}^{-1/2}$ is the upstream bulk Lorentz factor in the downstream rest frame.  

\subsection{Results}
\begin{figure}
\centering
\plotone{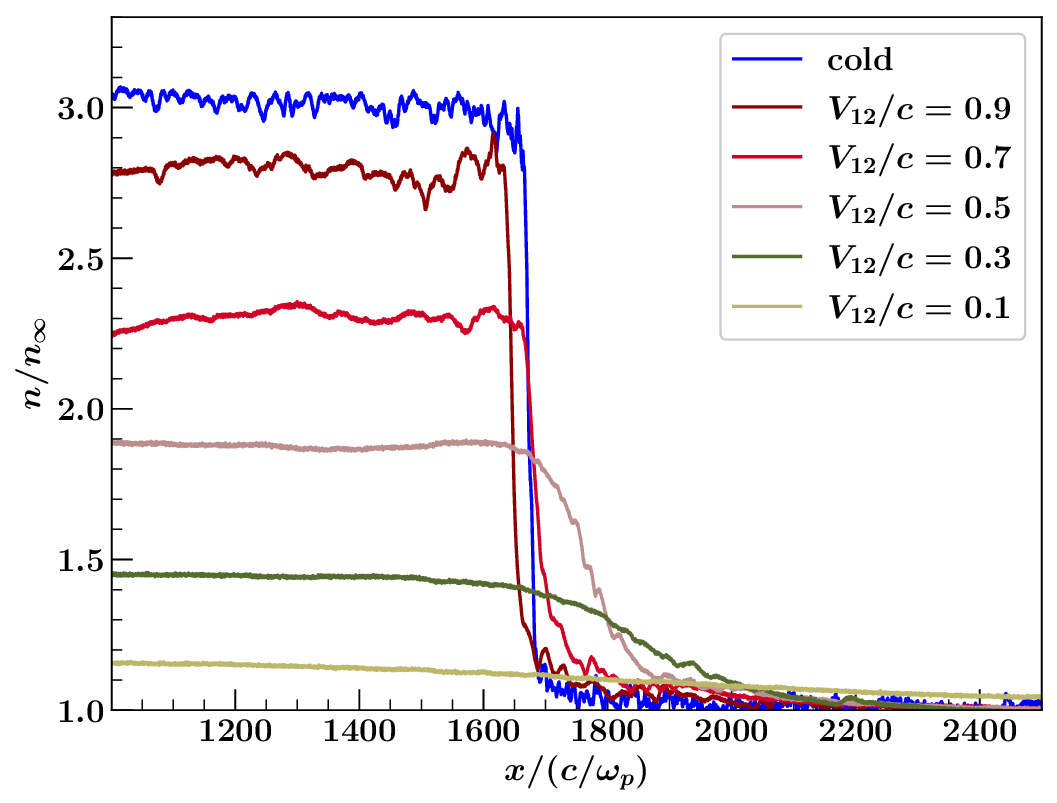}
\caption{One-dimensional density structure averaged over the y-direction in the downstream rest frame at $t=3600~\omega_{\rm p}^{-1}$. The density is normalized by the upstream density. 
The blue line shows the result for the cold upstream plasma ($kT/mc^2 = 0.01, \Gamma_{12}=24.57$). The other lines show results for the relativistically hot plasma ($kT/mc^2 = 5$).} 
\label{fig2}
\end{figure}
\begin{figure}
\centering
\plotone{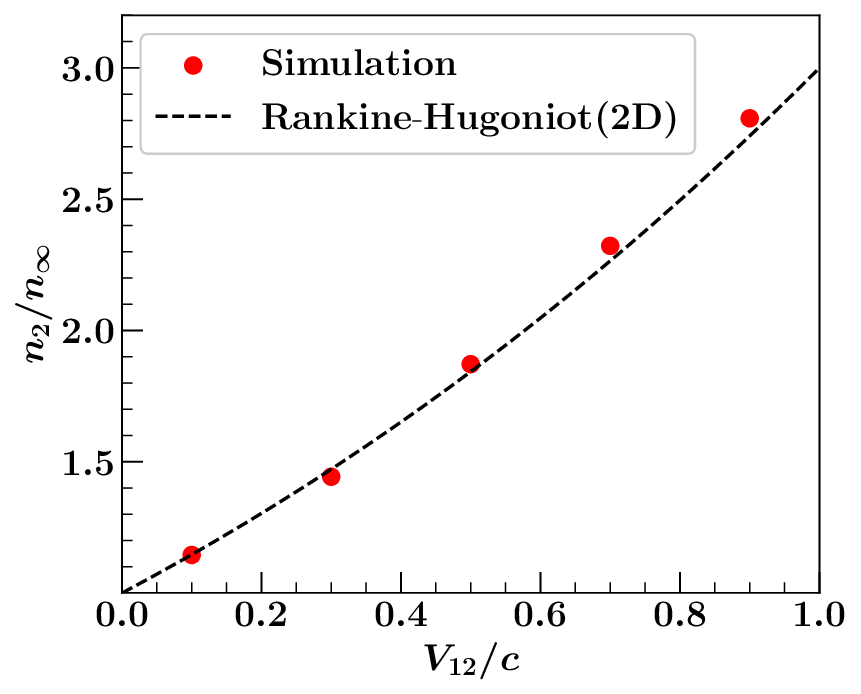}
\caption{Ratio of the downstream density to the upstream density in the downstream rest frame. The red points show the simulation results for the relativistically hot plasma ($kT/mc^2 = 5$). The dashed line shows the Rankine-Hugoniot relation in two-dimensional system.}
\label{fig3}
\end{figure}
\begin{figure*}
\centering
\includegraphics[scale=0.33]{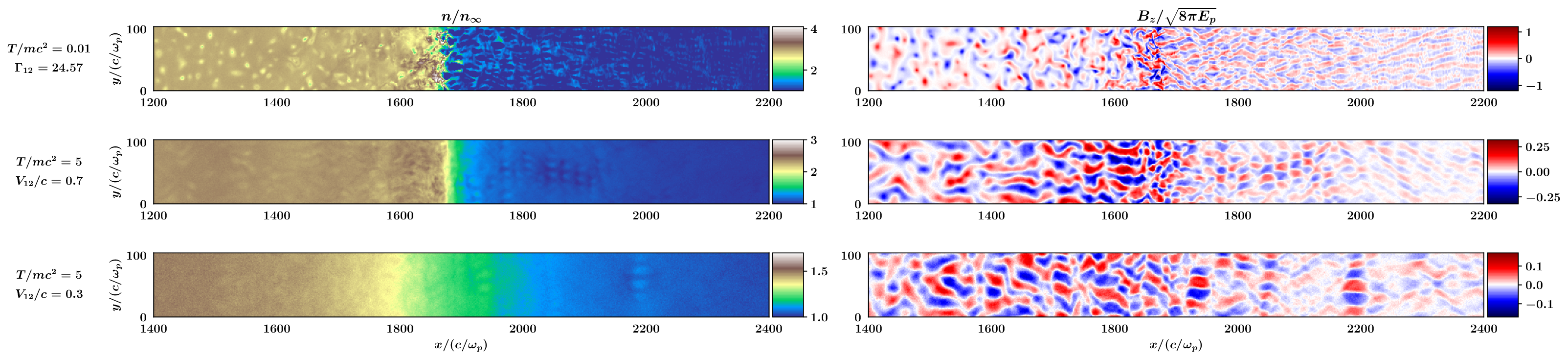}
\caption{Two dimensional distribution of the electron-positron density (left) and the $z$ component of the magnetic field (right) at $t=3600~\omega_{\rm p}^{-1}$. 
The top, middle, and bottom panels show simulation results for ($kT/mc^2 = 0.01, \Gamma_{12}=24.57$), ($kT/mc^2 = 5, V_{12}/c=0.7$), and ($kT/mc^2 = 5, V_{12}/c=0.3$), respectively. }
\label{fig4}
\end{figure*}
\begin{figure}
\centering
\plotone{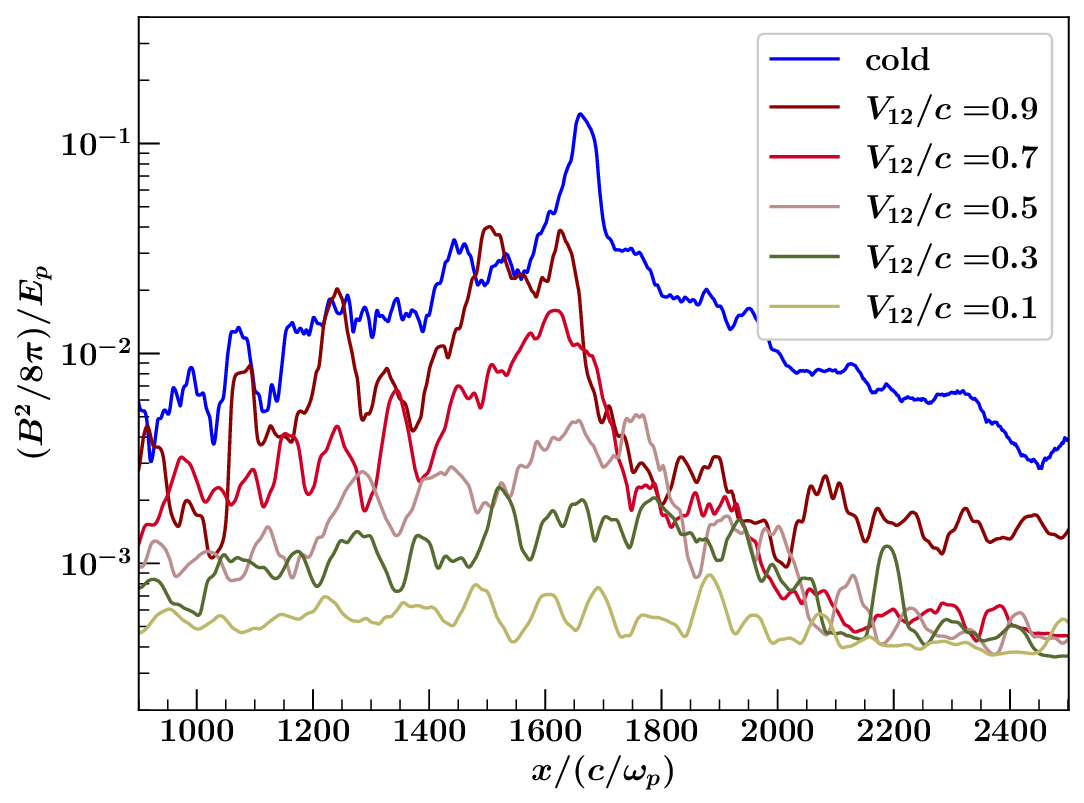}
\caption{The same as Figure~\ref{fig3}, but for the magnetic field energy density normalized by the upstream particle energy.}
\label{fig5}
\end{figure}
\begin{figure}
\centering
\plotone{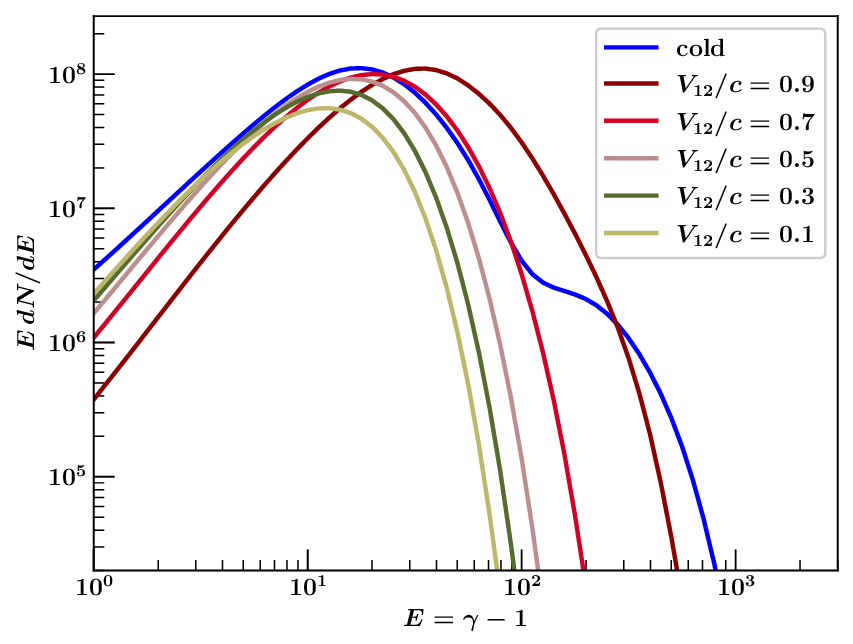}
\caption{Energy spectra of the downstream particles in the region of $800 < x / ( c/\omega_p) < 1600$ at $t=3600~\omega_{\rm p}^{-1}$.}
\label{fig6}
\end{figure}

All simulation results shown here are at $t=3600~\omega_{\rm p}^{-1}$. 
Figure~\ref{fig2} shows the one-dimensional electron-positron density structure averaged over the y-direction in the downstream rest frame  (simulation frame). 
Shock like structures are formed not only in the cold upstream plasma, but also in the relativistically hot upstream plasma for all runs. 
The spatial scales of the shock like transition layer do not significantly depend on the upstream flow velocity for $V_{12}/c > 0.5,~(M>1.2)$, but become longer with decreasing the Mach number. 

Figure~\ref{fig3} shows the relation between the upstream velocity at the downstream rest frame, $V_{12}/c$, and the density ratio across the shock like structure. 
The red points show the simulation results and the dashed line shows the Rankine-Hugoniot relation in the two-dimensional system. 
Simulation particles are scattered mainly in the $x$-$y$ plane because only the $z$ component of the magnetic field is strongly generated in the two-dimensional PIC simulation. 
Hence, although we solve the all three components of the equation of motion ($d\vec{p}/dt$), the simulation system can be regarded as the two-dimensional momentum system. 
The simulation results are in good agreement with the Rankine-Hugoniot relation, indicating that collisionless shocks are fully generated in our PIC simulations within $t\sim 10^3~\omega_{\rm p}^{-1}$ even in the relativistically hot unmagnetized electron-positron plasma.

To understand what plasma instabilities occur around the collisionless shock front in the relativistically hot upstream plasma, 
we show two-dimensional distributions of the density (left) and magnetic field (right) in Figure~\ref{fig4}. 
The top, middle, and bottom panels show results for the cold, hot plasma cases with $V_{12}/c=0.7$ and $0.3$, respectively.  
The density and magnetic field are normalized by the far upstream density, $n_{\infty}$, and $\sqrt{8\pi E_{\rm p}}$, respectively, where $E_{\rm p}=\int mc^2\sqrt{(u/c)^2+1}  f(u_x,u_y,u_z)d^3u$ is the upstream particle energy density and $f(u_x,u_y,u_z)$ is the distribution function in the four velocity space, $(u_x,u_y,u_z)$. 
For the Maxwell-J\"{u}ttner distribution with the drift velocity of $V_{12}$, $E_{\rm p}$ is represented by
\begin{equation}
E_{\rm p} = \Gamma_{12} n_{\infty}mc^2 \left(\frac{K_1(mc^2/kT)}{K_2(mc^2/kT)} + \frac{3+(V_{12}/c)^2}{mc^2/kT}\right) ,
\end{equation}
where $K_1(x)$ and $K_2(x)$ are the modified Bessel functions of the first and second kind, respectively.
For the cold limit ($kT\ll mc^2$), $E_{\rm p}$ is reduced to 
\begin{equation}
E_{\rm p} = \Gamma_{12} n_{\infty}mc^2 ,
\end{equation}

For all cases, the $z$ component of the magnetic field is much stronger than the other components of electromagnetic fields. 
All the patterns for hot plasma cases are similar to those for the cold upstream plasma case, but the characteristic length scale is longer for the hot upstream plasma cases and it is about $10~c/\omega_{\rm p}-30~c/\omega_{\rm p}$. 
The Weibel instability can explain these density and electromagnetic properties. 
The temperature anisotropy exists in the shock transition layer. 
In addition, the upstream plasma is unmagnetized. 
In such a system, it is inconceivable that a magnetic field could be generated by anything other than the Weibel instability.
The wavelength of the most unstable Weibel mode is longer for a smaller temperature anisotropy \citep{yoon07}. 
Since the temperature anisotropy in the shock transition region is smaller for a slower shock velocity and in the hot upstream plasma, our simulation results shown in Figure~\ref{fig4} are consistent with the characteristic of the Weibel instability.

As one can see in Figure~\ref{fig4}, when the shock velocity is slower and the upstream plasma is relativistically hot, the decay of the downstream magnetic field appears to be slower. 
To investigate the decay of the downstream magnetic field on a larger scale, we plot the one-dimensional distribution of the magnetic field energy density averaged over the y direction in Figure~\ref{fig5}, where the all curves are smoothed in $20~c/\omega_{\rm p}$ to remove small scale structures. 
The magnetic field energy density is normalized by the upstream particle energy, $E_{\rm p}$. 
Around the shock front ($x\sim 1750~c/\omega_{\rm p}$), the normalized magnetic field strength depends on the shock velocity and whether the upstream is cold or hot. 
The decay of the magnetic field is slower for slower shock velocities and the relativistically hot upstream plasma cases, 
so that the dependence is very weak in the downstream region ($x\sim 1000~c/\omega_{\rm p}$). 
Therefore, our simulations show that unmagnetized collisionless shocks in the relativistically hot plasma efficiently generate downstream magnetic fields even for weak shocks.  
This is because, as shown in Figure~\ref{fig4}, the characteristic length scale of the magnetic field is longer for the hot plasma cases and slower shock velocities, and magnetic fields generally decay more slowly at the longer length scale.

Finally, we show the energy spectra of downstream particles in Figure~\ref{fig6}. 
It has been shown that unmagnetized relativistic collisionless shocks in the cold plasma accelerate particles \citep{spitkovsky08b,martins09}. 
We confirmed generation of the nonthermal component for the shock in the cold upstream plasma (blue curve).  
However, for shocks in the relativistically hot upstream plasma, the nonthermal component does not appear in our simulations (other curves). 
The acceleration time is longer for slower shock velocities \citep{ohira19}, which may be the reason why the nonthermal component does not appear in shocks in the relativistically hot plasma. 
We need to perform a longer simulation to address this problem.

\section{Application to gamma-ray bursts}
\label{sec4}
\begin{figure}
\centering
\plotone{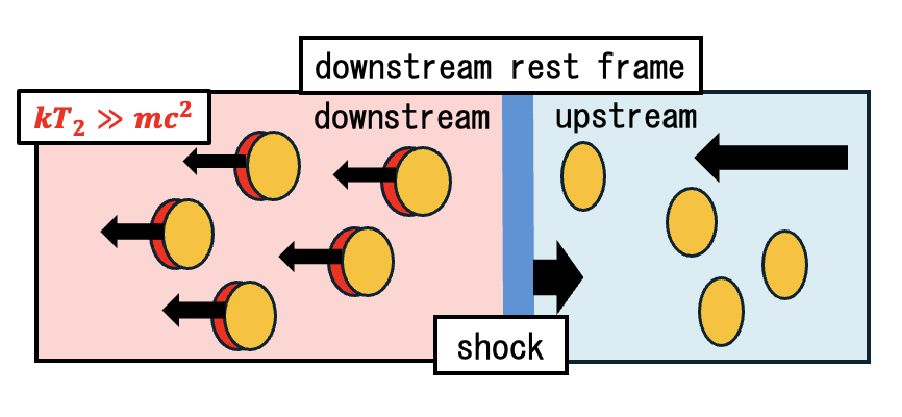}
\caption{Schematic picture of a relativistic shock propagating in an inhomogeneous density medium in the shock downstream rest frame. The vertical blue solid line shows the main relativistic shock. 
The average density regions (pink) have the average velocity of zero and a relativistic temperature in the downstream region. 
The high-density regions do not completely decelerated at the shock and move toward the far downstream.  
Then, the high-density regions push the front plasma in the downstream region of main shock, so that shocks start to propagate in the relativistically hot plasma (red).}
\label{fig7}
\end{figure}
An application of this work to GRBs is discussed in this section.
To explain prompt and afterglow emissions from GRBs, the internal and external shock models \citep{rees94,sari98} require the presence of accelerated electrons in the downstream region and sufficiently strong downstream magnetic fields to make synchrotron radiation efficient. 
The strong magnetic field in the internal shock model could originate from the base of the relativistic outflow. 
Because the external shock propagates in the interstellar or circumstellar media, the shock compressed magnetic field can be estimated, which is smaller than the strong magnetic field required in the standard external shock model \citep{santana14}. 
Thus, the external shock needs some magnetic field generations around the external shock or in its downstream emission region. 
The magnetic field generation may also work even in the internal shock. 
Some mechanisms of the magnetic field generation or amplification in GRBs have been proposed in the kinetic \citep{spitkovsky08a,chang08,keshet09,tomita16,tomita19,groselj24} and magnetohydrodynamical scales \citep{sironi07,inoue11,mizuno11,tomita22,morikawa24}.

We propose another model of the kinetic magnetic field generation as an application of this study. 
The schematic picture of our idea is shown in Figure~\ref{fig7}.
We consider a relativistic shock propagating in an inhomogeneous density medium. 
In the downstream rest frame of the main shock, after passing through the main shock (blue line), the average density regions slow down, and have the average velocity of zero and a relativistic temperature in the downstream region. 
On the other hand, the velocity of high-density regions (orange) does not drop to zero immediately after passing through the main shock because of the high inertia or ram pressure. 
Then, the shocked high-density regions push the front plasma, so that shocks start to propagate in the relativistically hot plasma (red). 
Therefore, we can expect many shocks in the downstream of the main shock if there are many high-density regions. 
As shown in Section~\ref{sec3}, each downstream shock generates magnetic fields through the Weibel instability. 
As a result, on average, strong magnetic fields can be maintained in the large downstream region of the main shock even if magnetic fields generated at the main shock decay quickly.

In addition to the magnetic field generation, the downstream shock would be important for particle acceleration.  
The interaction between high-density regions and the main shock drives the downstream turbulence, so that the downstream magnetic field can be amplified by turbulent dynamo \citep{sironi07,inoue11,mizuno11,tomita22,morikawa24}. 
It takes about the eddy turnover time to amplify the magnetic field by turbulent, that is, the turbulent dynamo cannot make a magnetic field turbulence just behind the main shock.  
Then, low-energy particles with a small gyroradius cannot be injected to particle acceleration by the main shock \citep{morikawa24}. 
On the other hand, our mechanism proposed in this section can quickly generate the magnetic field turbulence in the kinetic scale by the Weibel instability, so that the low-energy particles could be easily injected to particle acceleration by the main shock. 
Furthermore, the downstream shocks might directly accelerate particles through the diffusive shock acceleration \citep{axford77,krymsky77,bell78,blandford78}. 
Although no particles are accelerated in our simulation (see Figure.~\ref{fig6}). 
We need to perform a longer simulation to investigate the particle acceleration at the downstream shock, which will be addressed in future.

\section{Discussion and Summary}
\label{sec5}
In this work, we have for the first time investigated kinetic properties of collisionless shocks in a relativistically hot unmagnetized plasma. 
First of all, we showed by using the Rankine-Hugoniot relation that the dissipation fraction at shocks in the relativistically hot plasma is sufficiently large to explain the GRB observation even when the shock is not highly relativistic. 
Then, we have performed two-dimensional PIC simulations of collisionless shocks propagating in the relativistically hot ($kT=5mc^2$) unmagnetized electron-positron plasma. 
Our simulation results (Section~\ref{sec3}) are summarized as follows:
\begin{enumerate}
\item Weibel mediated collisionless shocks that satisfy the Rankine-Hugoniot relation are generated in the simulation time of $t = 3600~\omega_{\rm p}^{-1}$. 
\item The size of the shock transition region increases as the shock velocity decreases.
\item The downstream decay of the magnetic field generated by the Weibel instability around the shock is slower for slower shock velocities in the relativistically hot plasma cases. 
\item  No particles are accelerated around the shock in the relativistically hot plasma in the simulation time of $t = 3600~\omega_{\rm p}^{-1}$. 
\end{enumerate}
In addition, we have proposed a new mechanism of the magnetic field generation in GRBs as an application of this work (Section~\ref{sec4}). 

In addition to GRBs, there are relativistically hot plasmas in active galactic nuclei, pulsar wind nebulae, and around black holes and neutron stars. 
In this work, the upstream plasma was set to an unmagnetized electron-positron plasma with the Maxwell-J\"{u}ttner distribution as a first step. 
In general, the upstream plasma could be a magnetized and electron-ion (or electron-positron-ion) plasma. 
Moreover, the distribution function could have a nonthermal high-energy component. 
Collisioness shocks in these relativistically hot plasmas should also be investigated in future. 
This will open a new window for relativistic plasma physics. 

\begin{acknowledgments}
Numerical computations were carried out on Cray XC50 at the Center for Computational Astrophysics, National Astronomical Observatory of Japan. Y.O. is supported by JSPS KAKENHI grants No. JP21H04487 and No. JP24H01805. 
\end{acknowledgments}

\end{document}